\documentclass[prd,showpacs,psfig]{revtex4}
\usepackage{bm}
\usepackage{psfig}
\begin{document}
\title{
Static spherisymmetric solutions, gravitational lensing and
perihelion precession in Einstein--Kalb--Ramond theory}
\author{Sayan Kar}
\email{sayan@cts.iitkgp.ernet.in}
\affiliation{Department of Physics and Centre for Theoretical Studies
\\
Indian Institute of Technology, Kharagpur 721 302, WB, India}
\author{Soumitra SenGupta}
\email{tpssg@mahendra.iacs.res.in}
\affiliation{ Department of Theoretical Physics, Indian Association for the Cultivation of Science, 
\\
Jadavpur, Kolkata 700 032, India}
\author{Saurabh Sur}
\email{saurabh@juphys.ernet.in}
\affiliation{Department of Physics, Jadavpur University,
Kolkata 700 032, India}
\author{\ } 
\begin{abstract}
Static, spherically symmetric 
solutions of the Einstein--Kalb--Ramond (KR) field equations are
obtained. Besides an earlier known exact solution, we also
find an approximate, asymptotically flat solution for which
the metric coefficients are obtained as an infinte series in
$\frac{1}{r}$. Subsequently, we study gravitational lensing 
and perihelion precession in these spacetimes 
and obtain explicit formulae which include corrections to these
effects in the presence of the KR field.   
\end{abstract}
\maketitle
\newcommand{\be}{\begin{equation}}
\newcommand{\ee}{\end{equation}}
\newcommand{\bea}{\begin{eqnarray}}
\newcommand{\eea}{\end{eqnarray}}

\section{Introduction}

Gravitational theories in a curved background spacetime with torsion has been 
an area of investigation for a long time. Torsion, which appears as an 
antisymmetric tensorial piece in the connection \cite{hehl}, is arguably an 
inescapable consequence when the
matter fields giving rise to spacetime curvature are possessed 
with spin \cite{sab}. Therefore, the simplest level of torsion theory 
can provide a classical background
for quantum matter fields. Starting from the original Einstein-Cartan (EC) 
theory, several papers have appeared  which explore the various consequences 
of the torsion field, its impact on gravitational and cosmological solutions of the general 
relativistic field equations and on the nature of torsion couplings to arbitrary spin
fields\cite{torsion}. In recent years, a lot of inspiration in this regard is provided by 
superstring theory. In fact, it has been shown that spacetime torsion can be 
identified \cite{pmssg}
with the strength of the massless antisymmetric second-rank tensor field, viz., the Kalb-Ramond (KR) field $B_{\mu \nu}$ \cite{kr}, appearing in the 
heterotic string spectrum \cite{gsw}. Thus, torsion is, in some sense, 
an inherent feature in the
low-energy effective string action.

Extensive studies have already been carried out regarding the coupling of torsion with
other spin fields, especially the electromagnetic field, where the well-known
problem of violation of $U(1)$ gauge-invariance \cite{ham} is explored.
In the recent work of Majumdar and SenGupta \cite{pmssg}, this problem is sorted
out using the Chern-Simons (CS) extension which is augmented with the KR field
strength $\partial_{[\mu} B_{\nu \lambda]}$ on account of anomaly-cancellation of the
corresponding quantum theory. Such a formalism can have some 
important implications
in explaining certain astrophysically/cosmologically observable phenomena 
such as
cosmic optical activity \cite{skpmas,skpmss}, fermion helicity flip \cite{ssgas}, etc.
which may provide direct tests for the viability of string theory. However, a
complete understanding of such phenomena do require appropriate knowledge of the
various possible solutions of the gravitational field equations in spacetimes with
torsion (or, equivalently the KR field). In particular, in this regard, we have
explored earlier the possible existence of solutions for the static spherical symmetric
gravitational field equations in vacuum \cite{ssgss1} as well as the cosmological
solutions \cite{ssgss2} in presence of spacetime torsion. Identifying the KR field
stength three-tensor as the {\it Hodge}-dual of the derivative of the pseudoscalar
{\it axion} $H$ of string spectrum, it has been shown in \cite{ssgss1} that spherical
symmetric solutions can actually exist for a suitable form of the torsion (i.e., KR)
field and can have interesting features like that of the spacetime admitting a `wormhole'
or a `naked singularity'. However, apart from these exact solutions there may exist other
solutions as well for spherical symmetric spacetimes with torsion.

In this paper, we try to carry out the most general study of the existence of possible
static spherical symmetric solutions of the vacuum field equations, especially to check
the uniqueness of those obtained in the earlier work \cite{ssgss1}. 
Our aim is to establish a physically meaningful general solution which match the limiting requirements and provide a proper understanding about how the 
standard Schwarzschild solution gets
modified in the presence of different forms of the torsion field. Implications 
thereof.
are obtained through the study of
geodesic motion in such spacetimes. We  also
investigate the corrections inflicted by torsion on some of the standard tests 
of general relativity theory. As has been done
in the earlier paper \cite{ssgss1}, here also we seek solutions 
satisfying the boundary requirement of asymptotic flatness.

Although the full low energy effective action of string theory includes the graviton, dilaton, axion as well as other fields which 
may arise out of different type of compactifications, here we attempt to focuss on the effects of the axion only. We 
investigate whether the presence of axion ( appearing as a dual field to the Kalb-Ramond induced torsion ) is perceptible 
at all through the various observational and solar system tests. This is in contrast with the volume of works on 
the dilaton-graviton system\cite{coplidsey} and our work here aims to fill a gap in the literature.

\section{Static spherically symmetric solutions in a Kalb-Ramond background}

Following the identification \cite{pmssg} of the totally skew-symmetric torsion tensor
with the modified KR field strength $H_{\mu\nu\lambda}$, the action for gauge-invariant
EC-KR coupling is given by
\be
S = \int~ d^{4}x ~\sqrt{-g}~\left[\frac{R (g)}{\kappa} ~-~ \frac 1 {12}
H_{\mu \nu \lambda} H^{\mu \nu \lambda} \right]
\ee
which has exact correspondence with that in the low-energy effective string theory.
$R(g)$ is the Ricci scalar curvature and $\kappa \sim (Planck~mass)^{- 2}$ is the
gravitational coupling constant. The modified KR field strength three-form ${\bf H}$ is
defined by the KR field strength plus the $U(1)$ electromagnetic CS three-form: ~${\bf H}
~=~ d {\bf B} ~+~ \sqrt{\kappa}~ {\bf A \wedge F}$. Due to the Planck mass
suppression we neglect the CS term in the present analysis. The field equations that
can be obtained from the above action are given as
\bea
R_{\mu \nu} ~-~ \frac{1}{2} g_{\mu \nu} R ~=~ \kappa
{\cal T}_{\mu \nu} \\
D_{\mu} H^{\mu \nu \lambda} ~\equiv~ \frac{1}{\sqrt{- g}} \partial _{\mu}
(\sqrt{- g} H^{\mu \nu \lambda}) ~=~ 0
\eea
where $R_{\mu \nu}$ is the Ricci tensor of Riemannian geometry; and ${\cal T}_{\mu\nu}$
is a symmetric 2-tensor, analogous to the energy-momentum tensor, and is given by
\be
{\cal T}_{\mu \nu} ~=~  \frac 1 4 \left( 3 g_{\nu \rho} H_{\alpha \beta \mu}
H^{\alpha \beta \rho} ~-~ \frac{1}{2} g_{\mu \nu} H_{\alpha \beta \gamma}
H^{\alpha \beta \gamma} \right).
\ee
Taking the line element in its most general spherically symmetric form
\be
ds^{2} = e^{\nu (r,t)} dt^{2} - e^{\lambda (r,t)} dr^{2} - r^{2} (d \theta^{2}
+ \sin ^{2} \theta d \phi^{2})
\ee
and expressing the three-form $H_{\mu \nu \lambda}$ in terms of its Hodge-dual
one-form --- a pseudovector --- with independent components $H_{012}, H_{013},
H_{023}$ and $H_{123}$; it has been shown in \cite{ssgss1} that static spherical
symmetric solutions, consistent with the basic requirement of asymptotic flatness,
are possible only when $H_{023} \neq 0$ and all other components vanish. This
corresponds to the situation that the dual pseudoscalar `axion' $H$ defined
through the relation
\be
H_{\mu \nu \lambda} ~=~ \epsilon_{\mu \nu \lambda}^{~~~~ \sigma}~\partial_{\sigma} H.
\ee
depends on the radial coordinate $r$ only. Denoting $H_{023} H^{023}$ by $[h(r)]^2$
the field equations can be expressed as
\bea
e^{- \lambda}\left (\frac{1}{r^{2}} - \frac{\lambda'}{r}\right ) - \frac{1}{r^{2}}
~&=&~ \bar{\kappa} h^{2} \\
e^{- \lambda}\left (\frac{1}{r^{2}} + \frac{\nu'}{r}\right ) - \frac{1}{r^{2}}
~&=&~ - \bar{\kappa} h^{2} \\
e^{- \lambda} \left (\nu'' + \frac{\nu'^{2}}{2} - \frac{\nu' \lambda'}{2} + \frac{\nu'
- \lambda'}{r} \right ) ~&=&~ 2 \bar{\kappa} h^{2} \\
\partial_{1} \left(r^{2}~ e^{\frac{\nu - \lambda} 2} ~ H'\right) ~=~ \partial_{1} \left
(r^{2} h ~e^{\nu / 2}\right) ~&=&~ 0.
\eea
where a prime stands for derivative with respect to $r$; and the constant $\bar{\kappa}
~=~ \frac 3 4 \kappa$. The last equation in the above set is obtained by using both
Eq.(3) and the Bianchi identity for the KR field, viz., ~$\epsilon^{\mu \nu \lambda
\sigma}~\partial_{\sigma} H_{\mu \nu \lambda} ~=~ 0$. The above equations can
readily be solved to obtain
\be
h(r) ~=~ H'(r)~ e^{- \lambda / 2} ~=~ \frac{b_{0}}{r^{2}}~ e^{- \nu / 2}
\ee
and
\bea
e^{- \lambda} = 1 + \frac{c_{1}}{r} + \frac{\tau (r)}{r} \\
e^{\nu} = \frac{c_{2}}{r(r + \tau (r) + c_{1})} \exp \left[ \int^r ~ \frac{2 dr}{r
+ \tau (r) + c_{1}} \right]
\end{eqnarray}
where $b_0,~c_1$ and $c_2$ are the constants of the integrations and
\be
\tau (r) = ~\bar{\kappa} ~\int^r ~r^{2} h^{2} (r) dr.
\ee
The above solutions are consistent only when they satisfy the asymptotic flatness requirement,
viz., $e^{\pm\nu}, e^{\pm\lambda} \rightarrow 1$ as $r \rightarrow \infty$, and a consistency
condition derived from the field equations:
\be
\tau'' ~+~ \frac{\tau'}{r} ~=~ \frac{\tau' (\tau' - 1)}{r + c_{1} + \tau}.
\ee
The asymptotic flatness condition on the solutions demands that we must have $c_2 = 1$. This
can readily be verified in the limit where torsion vanishes, i.e., $\tau(r) = 0$. For
non-zero torsion, if we further put $c_1 = 0$, then as is shown in \cite{ssgss1}, a typical
exact solution satisfying the above requirement (15) can be obtained for a specific form
of $\tau(r)$, viz., $\tau (r) = - b/r ; ~~ b = \bar{\kappa} b_0^2 =$ constant (i.e.,
$h(r) \sim \frac 1 {r^2}$), whence
\bea
e^{- \lambda} &=& 1 - \frac{b}{r^{2}} \\
e^{\nu} &=& 1.
\eea
and we have a wormhole for a real KR field, i.e., a positive $b$. Note that
this geometry has been discussed many times in the literature beginning
with the work of Ellis {\cite{ellis}}, though its appearance in the
context of the Kalb--Ramond field coupled to gravity had not been
noticed till recently. 

\noindent
Now, to investigate whether the above solution is unique, or whether there 
exist other
{\it physically meaningful} solution(s) consistent with the boundary requirements, we
take a general functional form of $\tau (r)$, which depends on the KR field strength
$h(r)$, in the following section. Moreover, throughout our subsequent analysis we take
the KR field to be real.

Observing that ~$\tau(r)$~ does not involve any additive constant, we can generally
express it in the form
\be
\tau (r) ~=~ \sum_{m = 1}^{\infty} a_m r^m ~+~  \sum_{n = 1}^{\infty} \frac{b_n}{ r^n}.
\ee
But from Eqs.(11) and (14) we find
\be
\tau' (r) ~=~ \frac b {r^2} ~ e^{- \nu},~~~ b = \bar{\kappa} b_0^2.
\ee
Since $e^{- \nu} \equiv g^{00} \rightarrow 1$ as $r \rightarrow \infty$, Eqs.(18) and
(19) are consistent only when all the $a_m$'s in $\tau (r)$ vanish, i.e., $\tau (r) ~=~
\sum_{n = 1}^{\infty} b_n/r^n$. Plugging this in Eq.(15) and matching the coefficients
of equal powers of $r$ from both sides, we obtain
\bea
\tau (r) ~=~ b_1~ [~ \frac 1 r ~-~ \frac{c_1}{2 r^2} ~+~ \frac{c_1^2}{3 r^3}
&-& \left(1 - \frac{b_1}{6 c_1^2}\right) \frac{c_1^3}{4 r^4} \nonumber\\
&+& \left(1 - \frac{b_1}{2 c_1^2}\right) \frac{c_1^4}{5 r^5}
~-~ \left(1 - \frac{59 b_1}{60 c_1^2} + \frac{3 b_1^2}{80 c_1^4}\right)
\frac{c_1^5}{6 r^6} ~+~ \cdots~ ]
\eea
Computing $\tau' (r)$ and comparing with Eq.(19) we find $b_1 = - b$ and the solutions
can, in general, be expressed as
\bea
e^{\nu} ~\equiv~ g_{00} (r)  ~&=&~ 1 ~+~ \frac{c_1} r ~+~ \frac {b c_1}{6 r^3}
~-~ \frac {b c_1^2}{6 r^4} ~+~ \frac {6 b c_1^3 ~+~ 3 b^2 c_1}{40 r^5} ~+~ \cdots \\
e^{- \lambda} ~\equiv~ - g^{11} (r) ~&=&~ 1 ~+~ \frac{c_1} r ~-~ \frac b {r^2}
~+~ \frac {b c_1}{2 r^3} ~-~ \frac {b c_1^2}{3 r^4} ~+~ \left(b c_1^3 ~+~
\frac{b^2 c_1} 6\right) \frac 1 {4 r^5} ~+~ \cdots
\eea
and the solution for the KR field is given by
\be
h (r) ~=~ \sqrt{\frac b {\bar{\kappa}}}~ \frac 1 {r^2} ~ \left[ 1 ~-~ \frac{c_1} r
~+~ \frac{c_1^2}{r^2} ~-~ \left(c_1^3 ~+~ \frac {b c_1} 6\right) \frac 1 {r^3} ~+~
\left(c_1^4 ~+~ \frac{b c_1^2} 2\right) \frac 1 {r^4} ~+~ \cdots \right].
\ee
The above solutions are, by construction, asymptotically flat and reproduce the exact
solution found in \cite{ssgss1} for $c_1 = 0$. For $c_1 \neq 0$, we obtain the standard
Schwarzschild solution, viz., $e^{\nu} = e^{- \lambda} = 1 - r_s/r$ in the zero-torsion
limit, i.e, $b = 0$, provided $- c_1 = r_s = 2 G M$ (the Schwarzschild radius). In
fact, whenever $c_1$ is non-vanishing, being a constant we can always identify it with
$- r_s$, thereby obtaining the Schwarzschild solution in the limit $b \rightarrow 0$.

\section{Geodesics,lensing and perihelion precession in KR background}

The equations of geodesics for the general static spherically symmetric 
metric is given
by \cite{ch,wein}:
\bea
\dot{r}^2 &\equiv& \left(\frac{dr}{d\tau}\right)^2 ~=~ e^{- \lambda (r)} ~\left[ e^{- \nu
(r)} E^2 ~-~ \frac{J^2}{r^2} ~-~ L \right] \\
\dot{\phi} &\equiv& \frac{d\phi}{d\tau} ~=~ \frac J {r^2} \\
\dot{t} &\equiv& \frac{dt}{d\tau} ~=~ E e^{- \nu (r)}
\eea
where the motion is as usual considered to be taking place in the ~$\theta = \pi/2$~
plane and the constants $E$ and $J$ having the respective interpretations of energy
per unit mass and angular momentum about an axis perpendicular to the invariant
plane ($\theta = \pi/2$). Here, $\tau$ is an affine parameter and $L$ is the Lagrangian
having the values $0$ and $1$ respectively for null and time-like particles. We are not
concerned about the space-like particles herein. Now, let us consider separately the
following two cases:

\subsection{Case : ~$c_1 = 0$}

In this case, the metric coefficients $e^{\nu}$ and $e^{\lambda}$ given by Eqs.(16)
and (17) yield the equations for the radial geodesics ($J = 0$) :
\be
\left(\frac{dr}{dt}\right)^2 =~ (1 ~-~ L/E^2) \left(1 ~-~ \frac b{r^2}\right) ;~~~~~~~
\frac{dt}{d\tau} = E.
\ee
with solution
\be
t ~=~ \pm \frac{\sqrt{r^2 ~-~ b}}{\sqrt{1 ~-~ L/E^2}} ~+~~ constant
\ee
and the affine parameter $\tau ~\propto~ t$. The above equation represents
a hyperbola and shows that to an external observer a radially infalling particle
(time-like or null) approaches the radius $r = \sqrt{b}$ asymptotically but can
never reach it. As $\tau$ is linear in $t$ we find that the $\tau - r$ relationship
also represents a hyperbola. Now, for time-like geodesics ($L = 1$), $\tau$ is the
proper time and hence an observer falling with a time-like particle also skirts
the physical singularity at $r=0$ by asymptotically grazing the critical radius
at $r = \sqrt{b}$. This feature is in sharp contrast with what happens in a
Schwarzschild spacetime and is the characteristic of a wormhole spacetime.

\noindent
For the general motion of geodesics in the present case, we obtain from Eqs.(24) and
(25) the equation of orbit
\be
\left(\frac{du}{d\phi}\right)^2 ~=~ \left(\frac{E^2 - L}{J^2} ~-~ u^2\right) (1 ~-~
b u^2).
\ee
where $u = 1/r$. Now, in order to have bound orbits ($E^2 < 1$) the equation
~$du/d\phi = 0$~ must have at least two real, positive roots not coinciding with
any physical or coordinate singularity. Such roots exclusively imply the two turning
points of the closed orbit. In the present case, the real positive values of $u$ for
which ~$du/d\phi$ vanishes are $1/\sqrt{b}$ and $E/J$ for null geodesics ($L = 0$).
However, as the metric diverges at $u \equiv 1/r = 1/\sqrt{b}$ we infer that the
null geodesics cannot follow closed orbits in a wormhole spacetime. For time-like
geodesics ($L = 1$) in such a spacetime, there is no positive real value of $u$ for
which the metric is non-singular and $du/d\phi= 0$. Therefore, bound orbits are
not permissible for time-like geodesics also, i.e., for all kinds of particles we can
only have unbound orbis ($E^2 > 1$). A close inspection of  Eq.(29) raised to the
second order, viz.,
\be
\frac{d^2 u}{d \phi^2} ~+~ \left[ 1 ~+~ \frac b {D^2} \right] u ~=~ 2 b u^3 ;~~~~~~
D = \frac J {\sqrt{E^2 - L}}
\ee
shows that the KR field, represented by the parameter $b$, inflicts a two-fold change
in the $u - \phi$ straight line $u \sim \sin(\phi - \phi_{\infty})$ that corresponds to
the limiting Minkowski spacetime ($b = 0$). The KR field not only alters the intercept
on the $\phi$-axis but also produces a departure from the straight line motion due to
the term $b u^3$ on the right of the above equation. Although the impact parameter
$D$ and hence the changes in the intercept are different for massive and massless
particles, the amount of bending near the origin of force is same for both kinds of
particles. At $r = r_0$, the distance of closest approach towards the origin of force
$du/d\phi = 0$, whence Eq.(29) gives $r_0 = D$. Replacing back $u$ by $1/r$ and
solving Eq.(29) we obtain
\be
\phi (r) ~-~ \phi_{\infty} ~=~ \sin^{-1} (r_0/r) ~+~ \frac b {4 r_0^2} \left[ \sin^{-1}
(r_0/r) ~-~ (r_0/r) \sqrt{1 - (r_0/r)^2} \right] ~+~ O\left(\frac b {r_0^2}\right)^2.
\ee
The angle of bending for all types of particles is given by
\be
\Delta \phi ~=~ 2 \mid\phi(r_0) - \phi_{\infty}\mid ~-~ \pi ~=~ \frac b {r_0^2} \frac{\pi}
4 ~+~ O\left(\frac {b^2}{r_0^4}\right).
\ee
Now, instead of finding an expression for $\Delta \phi$ as an expansion in powers of $\frac{b}{r_0^2}
 \equiv x$(say), one can also perform an exact integration and obtain the
following expression for the amount of bending :
\be
\Delta \phi ~=~ 2\left ( K\left [\frac{b}{r_0^2}\right ] \right ) ~-~ \pi
\ee
where $K[x]$ is the complete elliptic integral of the first kind. A plot
of $\Delta \phi$ as a function of $x$ shows a linear
region for small values of $b << r_0^2$ with a slope reasonably 
close to $\pi/4$ -- a fact which is demonstrated in the approximate
calculation of the bending angle discussed above.

\vspace{.4in}

\psfig{figure=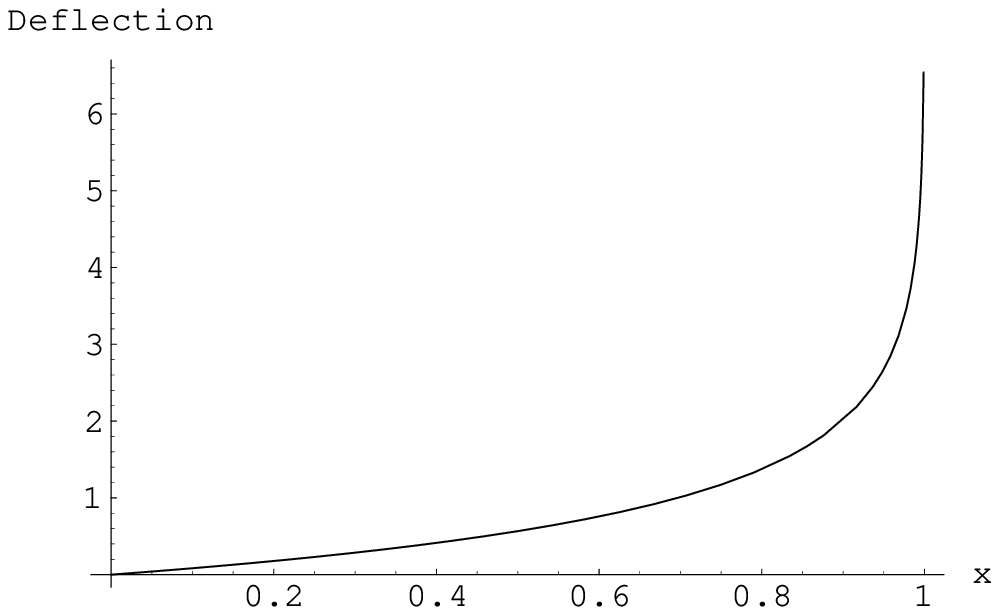,width=5.5in}

\vspace{.4in}
\noindent
Let us now turn towards the radial equation for  $u(\phi)$ quoted above. 
In order to obtain information about the trajectories of photons in the
geometrical optics limit we can, alternatively, solve for $u(\phi)$, in 
terms of elliptic functions by directly integrating the equation
given below.
\be
\left [ \frac{du}{d\phi} \right ]^2 ~=~ {b} \left (
\frac{1}{b} - u^2 \right ) \left (\frac{1}{r_0^2} - u^2 \right )
\ee
For $r_0^2=b$ the integration is straightforward and yields :
\be
r(\phi) ~=~ \frac 1 {u(\phi)} ~=~ \sqrt{b} ~coth (\phi -\phi_0)
\ee
In terms of the proper radial distance $l= \pm \sqrt{r^2 - b}$
we have:
\be
l ~=~ \pm \sqrt{b}~ cosech (\phi - \phi_0)
\ee
The general solution for $u(\phi)$ is given as :
\be
(\phi - \phi_0) ~=~ \frac{\sqrt{b}}{r_0} F\left [\arcsin{r_0 u}, \frac{b}
{r_0^2}\right ]
\ee
where $F$ denotes the incomplete elliptic integral of the first kind. 
On inverting we obtain :
\be
u (\phi) ~=~ \frac{1}{r_0} sn \left [\frac{r_0}{\sqrt{b}}(\phi-\phi_0), \frac{
b}{r_0^2} \right ]
\ee
where $sn$ denotes the Jacobian elliptic function. A plot of 
$r_0 u$ versus $\phi$ with $\phi_0 = 0$ and $\frac{r_0}{\sqrt{b}} = 3$
is shown below.

\vspace{.4in} 

\psfig{figure=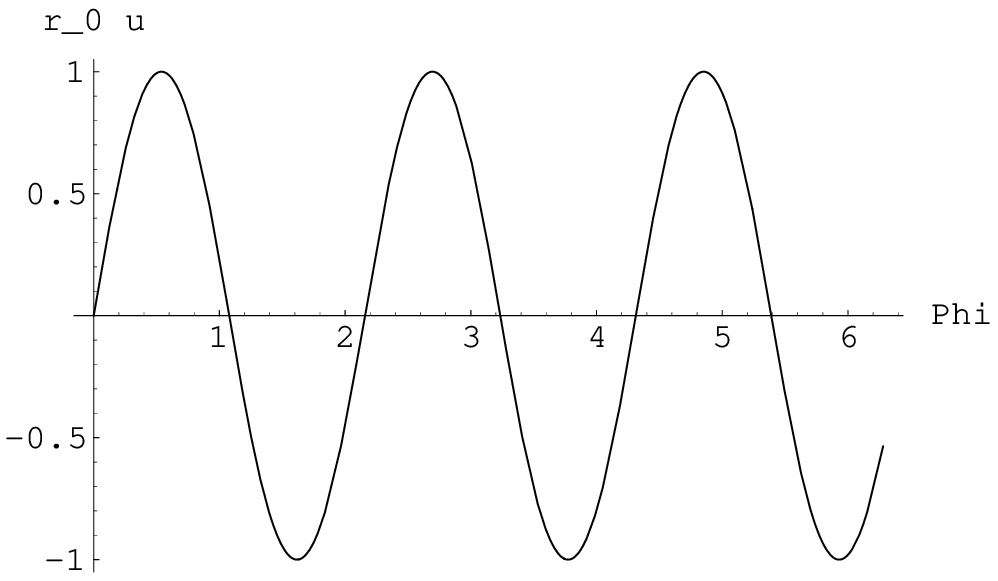,width=5.5in}

\vspace{.4in} 
\noindent
For $r_0=\sqrt{b}$ it can be seen that the $sn$ reduces to the hyperbolic
tangent which yields the expression discussed above for this case.
As $r_0\rightarrow \sqrt{b} $ the $\Delta \phi$ shoots up rapidly. The
trajectory exhibits multiple winding in the vicinity of the throat.
Why does this happen? Does the negative energy density present in the
vicinity of the throat result in such peculiar behaviour? It is
also clear that for other values of the ratio $x = b/r_0^2$
($x<1$) we find that the negative energy source (here the Kalb--Ramond
field) deflects the light ray by an amount which is crucially 
dependent on the KR parameter $\sqrt{b}$.
The explicit behaviour of the null geodesics including the case
of multiple windings for $r_0 = \sqrt{b}$ is also exhibited in the functional
forms for the trajectories derived above.

\subsection{Case : ~$c_1 \neq 0$}

Here we have the complete series solution given by Eqs.(21) and (22) with $c_1$
identified as $- r_s$, the Schwarzschild radius. Since both $e^{\nu}$ and $e^{\lambda}$
are convergent in the region $r >> r_s$, we study the motion of geodesics in such a
region where for simplicity, we consider the torsion  to be small, i.e., $b/r^2 << 1$.
This implies that we are assuming that the small torsion (or, equivalently the KR field)
does not completely change the nature of the trajectory of particles as in the case $c_1
= 0$. It rather inflics a correction over the general relativistic phenomena like bending
of light and the perihelic precession of planetary orbits. Dropping terms of order cubic
or more in $r_s/r$ and $b/r^2$ we can approximately write the solutions (21) and (22) as
\bea
e^{\nu} &=& 1 ~-~ \frac{r_s} r \\
e^{- \lambda} &=& 1 ~-~ \frac{r_s} r ~-~ \frac b {r^2}.
\eea
Indeed, an analysis similar to that in the case $c_1 = 0$ shows that for time-like particles
closed orbits are really possible for the above truncated series form of the metric coefficients.

\subsubsection{Bending of light rays (lensing)}

For photons the trajectory equations (24) and (25) yield
\be
\left(\frac{dr}{d\phi}\right)^2 ~=~ r^4 e^{- \lambda(r)} \left(\frac{e^{- \nu(r)}}{D^2}
~-~ \frac 1 {r^2}\right) ;~~~~~~~~~~ D = \frac J E
\ee
with solution in the form of a quadrature:
\be
\phi(r) ~-~ \phi_{\infty} ~=~ \int_r^{\infty} \frac{e^{\lambda(r)/2}~ dr/r}{\sqrt{
e^{- \nu(r)} \frac{r^2}{D^2} ~-~ 1}}
\ee
At the distance of closest approach ($r_0$) to the center of force,
$\frac{dr}{d\phi}\mid_{r = r_0} = 0$, whence Eq.(41) gives $D^2 = r_0^2
e^{- \nu(r_0)}$. Plugging this in Eq.(42) and using the specific expressions
for the metric components given in Eqs.(39) and (40), we obtain
\bea
\phi (r) &=& \phi_{\infty} ~+~ \sin^{-1} (r_0/r) ~+~ \frac{r_s}{2 r_0} \left(2 -
\sqrt{1 - (r_0/r)^2} - \sqrt{\frac{r - r_0}{r + r_0}} \right) \nonumber \\
&+& \frac{3 r_s^2}{8 r_0^2} \left[\frac 1 2 \sin^{-1} \left(\frac{r_0} r\right)
- 2 \cos^{-1} \left(\frac{r_0} r\right) - \frac{r_0}{2 r} \sqrt{1 - \left(\frac{r_0} r\right)^2}
+ \frac 3 2 \sqrt{\frac{r - r_0}{r + r_0}} - \frac 1 6 \left(\frac{r - r_0}{r + r_0}\right)^{\frac
3 2}\right] \nonumber \\
&+& \frac b {4 r_0^2} \left[ \sin^{-1} (r_0/r) ~-~ (r_0/r) \sqrt{1 - (r_0/r)^2} \right]
~+~~ higher ~order ~terms.
\eea
The angle of bending is given by
\bea
\Delta \phi ~=~ 2 \mid\phi(r_0) - \phi_{\infty}\mid ~-~ \pi ~=~ \frac{2 r_s}{r_0}
&+& \frac{3\pi}{16} \left(\frac{r_s}{r_0}\right)^2 ~+~\frac b {r_0^2} \frac{\pi} 4
\nonumber \\
&+& higher ~order ~corrections.
\eea
Looking at the above expression we note that the first and third terms are
relevant as long as we are interested in results valid upto first order in the
mass $M ~(r_s \sim GM)$ and
the KR parameter $b$. The first term is ofcourse the usual
Schwarzschild bending, whereas the third term comes from the KR field.
Using these two terms we may rewrite the total bending as follows :
\be
\Delta \phi ~=~ \left (\Delta\phi\right )_{Schw} \left [ 1 + \frac{\left (
\Delta \phi \right )_{KR}}{\left (\Delta \phi \right )_{Schw}} 
\right ]
\ee
Now the maximum KR energy density (note that this is negative)
is obtained by using the minimum value of $r$ (which is $r_0$, the
impact parameter). This energy density can be written as :
\be
\vert \rho^{max}_{KR}\vert   ~=~ \frac{c^4}{8\pi G} \frac{b}{r_0^4}
          ~=~ \frac{8}{3\pi} \frac{Mc^2}{V_0} 
\frac{\left (\Delta \phi\right)_{KR}}{\left (\Delta \phi\right )_{Schw}}
\ee
where $V_0$ is the volume $\frac{4}{3}\pi r_0^3$.

\noindent
We have calculated this energy density using the error bars for the
current deflection of light measurements for Sun \cite{will01}. 
The amount of energy
per unit volume is enormous resulting in an unacceptable ambient KR temperature much larger 
than that for CMBR. This suggests that $\left ( \Delta \phi
\right )_{KR}$ has to be far less than the value of these error bars
in order to give a reasonable KR field energy density and therefore 
remains undetectable within the present day experimental precision.

\subsubsection{Precession of perihelion of planetary orbits}

As has been mentioned earlier bound orbits are really possible for time-like
particles when $c_1 \neq 0$. In the case of elliptic planetary orbits, we obtain 
from Eqs.(24) and (25)
\be
\left(\frac{dr}{d\phi}\right)^2 ~=~ r^4 e^{- \lambda(r)} \left(\frac{e^{- \nu(r)}E^2
~-~ 1}{J^2} ~-~ \frac 1 {r^2}\right).
\ee
At perihelia and aphelia, $r = r_{\mp}$ and $\frac{dr}{d\phi}\mid_{r = r_{\pm}} = 0$,
whence the above equation yields
\be
\frac 1 {r_{\pm}^2} ~-~ \frac{e^{- \nu (r_{\pm})} E^2}{J^2} ~=~ - \frac 1 {J^2}
\ee
Solving for $E$ and $J$ we obtain the trajectory equation in the form of the
quadrature \cite{wein}:
\be
\phi(r) ~-~ \phi(r_{-}) ~=~ \int_{r_{-}}^r e^{\lambda(r)/2} Y^{-1/2}(r)~ \frac
{dr}{r^2}
\ee
where
\be
Y(r) ~=~ \frac{e^{\nu(r_+)}r_{-}^2 [ e^{\nu(r)} - e^{\nu(r_{-})} ] ~-~ e^{\nu(r_-)}r_{+}^2 
[ e^{\nu(r)} - e^{\nu(r_{+})} ]} {r_{+}^2 r_{-}^2 e^{\nu(r)} [ e^{\nu(r_{+})} - e^{\nu(r_{-})} ]} ~-~
\frac 1 {r^2}.
\ee
Simplifying the above expressions and carrying out the integration, we obtain for the 
metric in Eqs.(39) and (40) the amount of rotation of the orbit per revolution :
\bea
\Delta \phi ~=~ 2 \mid\phi(r_{+}) - \phi_(r_{-})\mid ~-~ 2 \pi &=& \left[ \frac{3 r_s}
l ~+~ \frac{3 r_s^2}{8 l^2} \left(18 + e^2 \right) ~+~ \frac b {l^2} \left(1 + \frac{e^2} 
2 \right) \right] \pi \nonumber \\
&+&~~~~~ higher ~order ~corrections.
\eea
Here $L$ and $e$ are respectively the semilatus rectum and eccentricity of the elliptic
orbit and are defined by ~$2/l = (1/r_{+} + 1/r_{-}),~~r_{\pm} = (1 \pm e) a,~ a$ being the
semimajor axis.

\noindent
An analysis similar to the preceding section shows that the maximum 
KR energy density that can be obtained by using the minimum value of $r$ (which
in this case is the perihelion distance $r_{-}$) is given by
\be
\vert \rho^{max}_{KR}\vert   ~=~ \frac{c^4}{8\pi G} \frac{b}{r_{-}^4}
	~\sim~ \frac{Mc^2}{V} 
\frac{\left (\Delta \phi\right)_{KR}}{\left (\Delta \phi\right )_{Schw}}
\ee
where we have considered the eccentricity $e$ to be fairly small so that
$r_{-} \sim l$, the semilatus rectum, and $V$ is the volume $\frac{4}{3}\pi l^3$
of the two-body system. Using the standard observational data for the perihelic
precessions of Mercury, Earth and Icarus \cite{will01} we have calculated 
$\vert \rho^{max}_{KR} \vert$ with $\left (\Delta \phi\right)_{KR}$ of the
order of error bars. Once again the energy density is found to be extremely high,
thus suggesting that $\left (\Delta \phi\right)_{KR}$ should be much
less than the present day experimental error bars.
\section{Conclusions}

In this paper we have extended the previous work by to two of us (SSG and SS)
on spherically symmetric static solutions of the Kalb--Ramond field
coupled to gravity. We have arrived at an approximate asymptotically
flat solution to the field equations. The `approximation' is necessitated
by the fact that, as far as we could see, the resulting equations are
not exactly solvable. This approximate solution has the nice feature that
it has a Schwarzschild piece and another term related to the KR field.
In a sense, therefore our approximate solution represents the exterior
gravitational field of a massive spherically symmetric body
{\em in the presence} of a Kalb--Ramond field. In order to arrive at
observable signatures, or {\em measure} the extent of the KR field
we have subsequently studied gravitational lensing and perihelion precession. 
For the exact
background (Ellis geometry) we have been able to find exact expressions
for the amount of bending. For the approximate solutions the amount of
bending and perihelion precession is approximate but the formulae 
look reasonable enough to
justify our approximation. We also make a suggestion about how the
bending/perihelion precession due to the KR field may be tested in observations.
Our findings seem to imply that if the amount of bending/perihelion
precession has to lie within the error bars of the observational
results for the Sun, we would end up with an abnormally large
KR energy density and, subsequently a huge ambient background temperature
(a cosmic KR background). This being absurd we conclude that the
bending/perihelion precession amount is far below the known error bars
and, therefore, cannot be detected by the present level of precision
in the solar system tests
employed to test GR. Therefore, alternative tests (besides those that
are related to cosmology) need to be thought out
in order to prove the existence of the KR field.   
Such possibilities have already been explored in the context of optical activity 
of the elctromagnetic radiation from distant galactic sources \cite{skpmas}, helicity flip of 
solar neutrino \cite{ssgas} and in other areas.
Future work along these lines 
is in progress and will be communicated in due course.

\section*{Acknowledgment}

This work is supported by Project grant no. 98/37/16/BRNS cell/676 from the Board
of Research in Nuclear Sciences, Department of Atomic Energy, Government of India,
and the Council of Scientific and Industrial Research, Government of India.

\end{document}